\documentstyle[12pt]{article}
\begin{document}

\def \d {{\rm d}}

\title{Continuous coordinates for all impulsive {\it pp}-waves}

\author{
J. Podolsk\'y\thanks{E--mail: {\tt podolsky@mbox.troja.mff.cuni.cz}}
,\  K. Vesel\'y
\\ \\ Department of Theoretical Physics,\\
Faculty of Mathematics and Physics, Charles University,\\
V Hole\v{s}ovi\v{c}k\'ach 2, 18000 Prague 8, Czech Republic.\\ }

\maketitle

\baselineskip=19pt

\begin{abstract}
We present a coordinate system for a general impulsive gravitational
{\it pp}-wave in vacuum in which the metric is explicitly continuous,
synchronous and ``transverse''.
Also, it is more appropriate for investigation of particle motions.
\end{abstract}

\vfil\noindent


\vfil
\eject

The metric of widely known class of vacuum {\it pp}-waves
\cite{KSMH}, plane-fronted gravitational waves with parallel
rays, can be written in the form
\begin{equation}
\d s^2=2\,\d\zeta \d\bar\zeta-2\,\d u\d r-(f+\bar f)\,\d u^2\ , \label{E1}
\end{equation}
where  $f(u,\zeta)$ is an arbitrary function of the retarded time
$u$ and the complex coordinate $\zeta$ spanning the plane wave surfaces.
Impulsive waves of this type can easily be
constructed by considering functions of the form
$f=f(\zeta)\delta(u)$ with $\delta(u)$ being the delta function.
Although this form is illustrative with the
pulse evidently localized along the null hyperplane $u=0$,
such a definition is only formal since the metric components contain
the delta function. There are several possibilities for introducing
the impulsive {\it pp}-waves correctly.

Naturally, an impulsive wave can be understood as a limit
of suitable sequence of sandwich waves.
This was done explicitly for $f=\zeta^2 d(u)$ using
various profiles $d(u)$, cf. \cite{Rindler}, \cite{PoVe}.

Of particular interest is the solution given by Aichelburg and Sexl
\cite{AicSex} in which  $f=\mu\log\zeta\>\delta(u)$ where $\mu$ is a real
constant. It was originally obtained by boosting a Schwarzschild
black hole to the speed of light while its mass is reduced to zero in an
appropriate way.  This solution describes an impulsive wave
generated by a single null particle at $\zeta=0$.
More general metrics have similarly been obtained from other
space-times of the Kerr-Newman family \cite{FerPen}-\cite{BaNa2}.

A general approach for constructing an arbitrary impulsive {\it pp}-wave
was proposed by Penrose \cite{Penrose}. His ``scissors-and-paste''
method is based on the removal of the null hyperplane $u=0$ from
Minkowski space-time and re-attaching the halves
${\cal M}_-(u<0)$ and  ${\cal M}_+(u>0)$ by making the
identification (a ``warp'') such that
$(\zeta,\bar\zeta,u=0,r)_{_{{\cal M}_-}}\equiv
(\zeta,\bar\zeta,u=0,
 r + {\textstyle \frac{1}{2}}[f(\zeta)+\bar f(\bar\zeta)])
_{_{{\cal M}_+}}$.

In the presented paper, however,  we adopt a different approach for rigorous
definition of a general impulsive (vacuum) {\it pp}-wave by writing
it in a simple coordinate system in which the metric components are explicitly
{\it continuous} for all values of $u$,
\begin{equation}
\d s^2=2 \left|\d\bar\eta-
 {\textstyle \frac{1}{2}}u\Theta(u) f''(\eta)\d\eta\right|^2 - 2\,\d u\d v\ ,
\label{E2}
\end{equation}
where $\Theta(u)$ is the Heaviside step function ($\Theta=0$ for
$u<0$, $\Theta=1$ for $u>0$) and $f'=\d f/\d\eta$.
For the metric (\ref{E2}) the delta function appears only in the
components of the curvature tensor.
The transformation relating (\ref{E2}) to (\ref{E1}) is
\begin{eqnarray}
\zeta&=&\eta-{\textstyle\frac{1}{2}}u\Theta(u) {\bar f}'  \ ,\nonumber\\
 r   &=&v   -{\textstyle\frac{1}{2}}\Theta(u)(f+\bar f)
            +{\textstyle\frac{1}{4}}u\Theta(u) f'{\bar f}' \ ,\label{E3}
\end{eqnarray}
where $f=f(\eta)$. Note that the transformation is discontinuous
at $u=0$ with the jump in $r$ given by
${\textstyle \frac{1}{2}}(f+\bar f)$
which corresponds to the Penrose identification method.

 From (\ref{E2}) it is
clear that for linear functions $f$  the metric
reduces  to Minkowski flat space-time. For other functions $f$
the metrics are non-trivial and describe impulsive
gravitational waves. In particular, when $f(\zeta)=C\zeta^n$
where $C=|C|\exp(in\varphi_n)$ is an arbitrary complex constant
and $n$ is an integer,
it is convenient to introduce two real ``polar'' coordinates
$\rho,\varphi$ by
$\eta={\textstyle\frac{1}{\sqrt2}}\rho\exp[i(\varphi-\varphi_n)]$.
In these coordinates the metric (\ref{E2}) takes the form
\begin{eqnarray}
\d s^2 &=&[1+u\Theta(u)D\rho^{n-2}]^2(\d\rho^2+\rho^2\d\varphi^2)
            \label{E4}\\
       && -4u\Theta(u)D\rho^{n-2}
             [\cos({\textstyle \frac{n}{2}}\varphi)\d\rho
             -\sin({\textstyle \frac{n}{2}}\varphi)\rho\d\varphi]^2
        - 2\,\d u\d v\ ,\nonumber
\end{eqnarray}
where $D=2^{-n/2}n(n-1)|C|$. We may also  assume coordinates
$x=\rho\cos\varphi$, $y=\rho\sin\varphi$,
\begin{eqnarray}
\d s^2&=& [1-2DA_n u\Theta(u)+D^2(x^2+y^2)^{n-2}u^2\Theta(u)]\,\d x^2
                             \nonumber\\
      &&  [1+2DA_n u\Theta(u)+D^2(x^2+y^2)^{n-2}u^2\Theta(u)]\,\d y^2
                             \label{E45}\\
      && + 4DB_n u\Theta(u)\, \d x\d y   - 2\,\d u\d v\ ,\nonumber
\end{eqnarray}
where
$A_n(x,y)={\cal R}e\,\{(x+iy)^{n-2}\}$,
$B_n(x,y)={\cal I}m\,\{(x+iy)^{n-2}\}$.

Again, for $n=0$ and $n=1$ the metric (\ref{E45}) reduces to Minkowski
space since $D=0$.
For $n=2$ we get $A_2=1$ and $B_2=0$ so that (\ref{E45}) gives
\begin{equation}
\d s^2= (1-u\Theta(u)|C|)^2\d x^2 + (1+u\Theta(u)|C|)^2\d y^2
        - 2\,\d u\d v\ ,
\label{E5}
\end{equation}
which is the well known continuous form of an impulsive
plane wave \cite{Penrose}.
Solutions (\ref{E4})  with $n>2$ are unbounded at $\rho=\infty$.
Alternatively, they may be written in the form  (\ref{E45}) with
$A_3=x$, $B_3=y$ for $n=3$,
$A_4=x^2-y^2$, $B_4=2xy$ for $n=4$ etc.

The  metric (\ref{E4}) is very suitable for describing impulsive
gravitational {\it pp}-waves generated by null particles
of an arbitrary multipole structure \cite{GrifPo} located at
$\rho=0$. These are
given by $n=-1, -2, \cdots$. Note that, interestingly, the Aichelburg
and Sexl ``monopole'' solution can also be written in the form
(\ref{E4}) with $n=0$ but $D=-4\mu\not=0$,
\begin{equation}
\d s^2= \left[1+\frac{4\mu}{\rho^2}u\Theta(u)\right]^2 \d\rho^2
       +\left[1-\frac{4\mu}{\rho^2}u\Theta(u)\right]^2 \rho^2\d\varphi^2
        - 2\,\d u\d v\ ,
\label{E6}
\end{equation}
which has been introduced in \cite{DEa}, \cite{DEaPa}.

Considering
$u={\textstyle\frac{1}{\sqrt2}}(t-z)$ and
$v={\textstyle\frac{1}{\sqrt2}}(t+z)$ we see that
the above coordinate systems are not only continuous but also
synchronous (gaussian). Moreover, they are explicitly
``transverse'', i.e., naturally adapted for describing gravitational
{\it pp}-waves propagating in the $z$-direction. In particular,
they are useful for a study of the motions of free particles
(avoiding some of the complications noticed in \cite{Ba}). Let us assume
particles in metric (\ref{E4}) standing at fixed $\rho_0, \varphi_0, z_0$ in
Minkowski (half) space $u<0$ ahead of the impulsive wave.
They move along geodesics with $t$ measuring their (synchronized)
proper times. After the passage of the impulse they
start to move one with respect to the other. In particular, the
{\it relative} distance between two nearby particles having the same
$\varphi_0, z_0$ but with radial coordinates differing by $\Delta\rho_0$, or
having the same $\rho_0, z_0$ but with a difference $\Delta\varphi_0$,
changes for small values
of $u={\textstyle\frac{1}{\sqrt2}}(t-z_0)>0$ according to
\begin{eqnarray}
\Delta l_\rho&=&\quad
\Delta\rho_0[1-{\textstyle\frac{D}{\sqrt2}}\rho_0^{n-2}\cos(n\varphi_0)(t-z_0)]
  \ ,\nonumber\\
\Delta l_\varphi&=&
\rho_0\Delta\varphi_0[1+{\textstyle\frac{D}{\sqrt2}}\rho_0^{n-2}\cos(n\varphi_0)(t-z_0)]\ .
\label{E8}
\end{eqnarray}
Therefore, when the particles approach in the radial direction
they  move apart in (perpendicular) tangential
direction and vice versa. Locally, the ring of free test
particles at any point  deforms into an ellipse similarly as is typical for
linearized gravitational waves. The effect vanishes (up to the
first order in $u$) in places where $\cos(n\varphi_0)=0$. For the
Aichelburg and Sexl solution ($n=0, D\not=0$) the above relative
motions do not depend on $\varphi_0$.
Similarly, the motions  are independent of $\rho_0$ for $n=2$, i.e.,
for plane gravitational wave (``homogeneous'' {\it pp}-wave).
For $n=3, 4, \cdots$ the effect given by (\ref{E8}) is more intense with a
growing value of $\rho_0$. On the other hand for solutions with
$n=0,-1,-2,\cdots$ representing impulsive waves generated by null
particles of an $n-$pole structure the relative motions of this
type vanish as $\rho_0\to\infty$.

\vspace{4mm}

We acknowledge the support of grants No. GACR-202/96/0206 and
No. GAUK-230/1996 from the Czech Republic and Charles University.


\vspace{2mm}


\begin{thebibliography}{99}
\setlength{\itemsep}{-1mm}

\bibitem{KSMH} D. Kramer, H. Stephani, M.A.H. MacCallum, E. Herlt,
    Exact Solutions of Einstein's Field Equations (Cambridge
    University Press, Cambridge, 1980) \S21.5.

\bibitem{Rindler}   W. Rindler,
    Essential Relativity (Springer, New York, 1977).

\bibitem{PoVe}  J. Podolsk\'y, K. Vesel\'y,
    Czech. J. Phys. (1998) in press.

\bibitem{AicSex} P.C. Aichelburg, R.U. Sexl,
    Gen. Rel. Grav.  2 (1971) 303.

\bibitem{FerPen} V. Ferrari, P. Pendenza,
    Gen. Rel. Grav.  22 (1990) 1105.

\bibitem{LoSa} C.O. Loust\'o, N. S\'anchez,
    Nucl. Phys. B,   383 (1992) 377.

\bibitem{BaNa1} H. Balasin, H. Nachbagauer,
    Class. Quantum Grav.  12 (1995) 707.

\bibitem{BaNa2} H. Balasin, H. Nachbagauer,
    Class. Quantum Grav.  13 (1996) 731.

\bibitem{Penrose} R. Penrose,  in:
    General Relativity, ed. L.O'Raifeartaigh
    (Clarendon Press, Oxford, 1972).

\bibitem{GrifPo}   J.B. Griffiths, J. Podolsk\'y,
    Phys. Lett. A  236  (1997) 8.

\bibitem{DEa}   P.D. D'Eath,
    Phys. Rev. D  18 (1978) 990.

\bibitem{DEaPa} P.D. D'Eath, P.N. Payne,
    Phys. Rev. D  46 (1993) 658.

\bibitem{Ba} H. Balasin,
    Class. Quantum Grav.  14  (1997) 455.


\end{thebibliography}
\end{document}